\documentclass[12pt]{article}
\usepackage{sc3conf}
\usepackage{amsfonts}
\usepackage{epsfig}

\newcommand{\sh}{\mathop{\mathrm{sh}}}
\newcommand{\be}{\begin{equation}}
\newcommand{\ee}{\end{equation}}
\newcommand{\rmd}{\mbox{d}}
\newcommand{\bea}{\begin{eqnarray}}
\newcommand{\eea}{\end{eqnarray}}

\begin{document}
\raggedbottom

\title{Integrals involving four Macdonald functions and their relation to $7\zeta(3)/2$}

\authors{ Cyril Furtlehner\adref{1} and St\'ephane Ouvry\adref{2}}

\addresses{\1ad  
INRIA, Preval, Domaine de Voluceau-Rocquencourt, B.P. 105,
78153 Le Chesnay Cedex, France
\nextaddress  
\2ad LPTMS, Bat 100, Universit\'e Paris Sud, 91405 Orsay, France
  }

\maketitle

\begin{abstract} A family of multiple integrals over four 
variables is rewritten in terms of a family of
simple integrals involving the product of four modified Bessel (Macdonald
functions). The latter are shown to be related to $7\zeta(3)/2$. A
generalization to $2n$ integration variables is given which yields only
$\zeta$ at odd arguments.

\end{abstract}

\section{Introduction}\label{sec:1}

Consider
\[ I_{1100}(\beta)=\int_0^{\beta}\rmd \beta_1\int_0^{\beta_1}\rmd \beta_2\int_0^{\beta_2}\rmd \beta_3
\int_0^{\beta_3}\rmd \beta_4{(a+c)(b+d)\over abc+bcd+cda+dab}
\]
where $a=\beta_1-\beta_2,b=\beta_2-\beta_3,c=\beta_3-\beta_4,
d=\beta+\beta_4-\beta_1$.
 One has
\[ I_{1100}(\beta)=\int_0^{\beta}\rmd a\,\int_0^{\beta-a}\rmd b\,\int_0^{\beta-a-b}\rmd c\, 
{d(a+c)(b+d)\over abc+bcd+cda+dab}\]
where $d=\beta-(a+b+c)$ is understood\footnote{same as 
\[ I_{1100}(\beta)=\int_{0\le a+b+c\le \beta} \rmd a\,\rmd b\,\,\rmd c\,
{d(a+c)(b+d)\over abc+bcd+cda+dab}\nonumber\]}.
By successive integration  on $c,b,a$ one has obtained \cite{1}
\[ I_{1100}(\beta)={1+{\tilde\zeta(3)}\over 16}\beta^3\]
where ${\tilde\zeta(3)}=7\zeta(3)/2$.
Rewrite $I_{1100}(\beta)$ differently
\[ I_{1100}(\beta)=\int_{a,b,c,d=0}^{\infty}\rmd a\, \rmd b\, \rmd c\, \rmd d\, {d(a+c)(b+d)\over abc+bcd+cda+dab}
\delta(\beta-(a+b+c+d))\]
and also
\[ I_{1100}(\beta)=\beta\int_{a,b,c,d=0}^{\infty}\rmd a\, \rmd b\, \rmd c\, \rmd d\, {ab\over abc+bcd+cda+dab}
\delta(\beta-(a+b+c+d))\]
Laplace transform
\[ I_{1100}(s)=\int_0^{\infty}\rmd \beta e^{-\beta s}{I_{1100}(\beta)\over\beta}=
{1\over s^3}\int_{a,b,c,d=0}^{\infty}\rmd a\, \rmd b\, \rmd c\, \rmd d\, {ab\over abc+bcd+cda+dab}
e^{-(a+b+c+d)}\]
such that $I_{1100}(s)=2 I_{1100}(\beta=1)/s^3$.
Fix $s=1$ and denote $I_{1100}(s=1)=I_{1100}$. One has
\[ I_{1100}=
 \int_{a,b,c,d=0}^{\infty}\rmd a\, \rmd b\, \rmd c\, \rmd d\, \int_0^{\infty}
\rmd t\, {1\over cd}
e^{-(a+b+c+d)-t({1\over a}+{1\over b}+{1\over c}+{1\over d})}\]
Change  variable $u=2\sqrt{t}$
\[ I_{1100}=2^4\int_{0}^{\infty}{u\rmd u\,\over 2}  ({u\over 2})^2K_1(u)^2K_0(u)^2\]
where the 
$K_{\nu}(u)$'s are   modified 
Bessel functions also called Macdonald functions \cite{2}.
Therefore
\be\label{resbis} I_{1100}=2^4\int_{0}^{\infty}{u\rmd u\,\over 2}  ({u\over 2})^2K_1(u)^2K_0(u)^2=
2{1+{\tilde\zeta(3)}\over 16}\ee


\section{Generalization} \label{sec:3}
To generalize (\ref{resbis}),
introduce for $n_a,n_b,n_c,n_d\in N$ 
\be\label{11}
I_{n_an_bn_cn_d}=\int_{a,b,c,d=0}^{\infty}\rmd a\, \rmd b\, \rmd c\, \rmd d\,
{a^{n_a}b^{n_b}c^{n_c}d^{n_d}\over abc+bcd+cda+dab} e^{-(a+b+c+d)}\ee
 and
rewrite it as\footnote{ This is also formally true for 
$n_a,n_b,n_c,n_d\in Z$ but (\ref{11}) diverges as soon as
$n_a<0$ or $n_b<0$ or $n_c<0$ or $n_d<0$. For $n_a,n_b,n_c,n_d\ge m-1, \quad m\ge 1$ 
\[\nonumber
I^m_{n_an_bn_cn_d}=\int_{a,b,c,d=0}^{\infty}\rmd a\, \rmd b\, \rmd c\, \rmd
d\, {a^{n_a}b^{n_b}c^{n_c}d^{n_d}\over (abc+bcd+cda+dab)^m} e^{-(a+b+c+d)}
\]
rewrites as 
\[
{1\over \Gamma(m)}2^4\int_{0}^{\infty}{u\rmd u\,\over 2}  
({u\over 2})
^{n_a+n_b+n_c+n_d-4(m-1)}
K_{n_a-(m-1)}(u)K_{n_b-(m-1)}(u)K_{n_c-(m-1)}(u)K_{n_d-(m-1)}(u)\nonumber
\]
The  considerations  that follow for  $I_{n_an_bn_cn_d}$ (i.e. the case $m=1$) 
generalizes to   $I^m_{n_an_bn_cn_d}$ for $m\ge 1$
implying that the latter  can also be built from $p_n(0000)$ and $p_n(1111)$ 
defined in (\ref{p0}) and (\ref{p1}).  
Therefore $I^m_{n_an_bn_cn_d}$ 
is of the form (\ref{conj}).
Note  that if one introduces back $s$ in  $I^m_{n_an_bn_cn_d}$
 \[ I^m_{n_an_bn_cn_d}(s)=\int_{a,b,c,d=0}^{\infty}\rmd a\, \rmd b\, \rmd c\, \rmd d\, {a^{n_a}b^{n_b}c^{n_c}d^{n_d}\over (abc+bcd+cda+dab)^m}
e^{-s(a+b+c+d)}\nonumber\]
and differentiates $n$ times with respect to $s$, one obtains with $s=1$
\[ I^{mn}_{n_an_bn_cn_d}=\int_{a,b,c,d=0}^{\infty}\rmd a\, \rmd b\, \rmd c\, \rmd d\, (a+b+c+d)^{n}{a^{n_a}b^{n_b}c^{n_c}d^{n_d}\over (abc+bcd+cda+dab)^m}
e^{-(a+b+c+d)}\nonumber\]
again of the form (\ref{conj}). 
} 
\[ I_{n_an_bn_cn_d}=2^4\int_{0}^{\infty}{u\rmd u\,\over 2}
({u\over 2})^{n_a+n_b+n_c+n_d} K_{n_a}(u)K_{n_b}(u)K_{n_c}(u)K_{n_d}(u)\]
 One has 
\be\label{conj}
I_{n_an_bn_cn_d}= {u_{n_an_bn_cn_d}+v_{n_an_bn_cn_d}{\tilde\zeta(3)}}\ee 
where
$u_{n_an_bn_cn_d}, v_{n_an_bn_cn_d}$ are positive or negative rational
numbers.

\noindent{\bf -Proof:}
Since 
\[ K_{n_a}(u)=2(n_a-1){
K_{n_a-1}(u)\over u}
+K_{n_a-2}(u)\]
then { if} $n_a>0$
even/odd
\bea {u^{n_a}\over 2^{n_a}}K_{n_a}(u)&=&
{u^{n_a}\over 2^{n_a}}K_{0/1}(u)+{u^{n_a-1}\over 2}K_{1/0}(u)+
c_{n_a-2}{u^{n_a-2}}K_{0/1}(u)+\ldots\nonumber\\
\nonumber
&+&c_1uK_{1}(u)\eea
where $c_{n_a-2},\ldots,c_1$ are rational numbers.
It follows that
 (\ref{conj}) is true if   
\be\label{p0} p_n(0000)=\int_{0}^{\infty}\rmd u\,  u^{n+1}
K_{0}(u)^4\quad n\ge 0\ee
\[ p_n(0011)=\int_{0}^{\infty}\rmd u\,  u^{n+1}
K_{0}(u)^2K_{1}(u)^2\quad n\ge 2\]
\be\label{p1} p_n(1111)=\int_{0}^{\infty}\rmd u\,  u^{n+1}
K_{1}(u)^4\quad n\ge 4\ee
($n$ even)

\noindent and 
\[ i_n(0001)=\int_{0}^{\infty}\rmd u\,  u^{n+1}
K_{0}(u)^3K_{1}(u)\quad n\ge 1\]
\[ i_n(0111)=\int_{0}^{\infty}\rmd u\,  u^{n+1}
K_{0}(u)K_{1}(u)^3\quad n\ge 3\]
($n$ odd)

\noindent are again\footnote{Note  that  a product of four Macdonald functions
is needed here: 
for example 
\[ \int_0^\infty {\rm d}u\, u K_0(u)^3 
={3\over 2}\sum_{p=0}^\infty \frac{1}{(3p+1)^2}
-\frac{2}{3}\zeta(2)\]
is not simply given in terms of $\zeta$'s.} of the form (\ref{conj}).
To show this, integrate by parts 
(use $\rmd K_{0}(u)/\rmd u\,=-K_{1}(u)$ and $\rmd
(uK_{1}(u))/\rmd u\,=-uK_{0}(u)$)
to obtain
\[ p_n(0000)={4\over n+2}i_{n+1}(0001) \quad n\ge 0\]
\[
p_n(0011)={2\over n}(i_{n+1}(0001)+i_{n+1}(0111))\quad n\ge 2\]
\[ p_n(1111)={4\over n-2}i_{n+1}(0111)\quad n\ge 4\]
which implies
\[ 2np_n(0011)=(n+2)p_n(0000)+(n-2)p_n(1111)\quad n\ge 4\]
It remains  to be found $p_n(0000)$ and $p_n(1111)$:  
integration by parts gives
\[ i_n(0111)={1\over n-1}((p_{n+1}(1111)+3p_{n+1}(0011))\quad n\ge 3\]
\[ i_n(0001)={1\over n+1}((p_{n+1}(0000)+3p_{n+1}(0011))\quad n\ge 1\]
thus the recurrence relation acting in a 2 dimensionnal vector space
\be\label{rec}
\left(
\begin{array}{c}
p_{n+2}(0000)\\ 
p_{n+2}(1111)\end{array}\right)
={1\over 2^5(n+2)}
\left(
\begin{array}{cc}
(n+2)^2(5n+4) & -3n^2(n-2)\\  
 -3(n+2)^2(n+4) & n(n-2)(5n+16) 
\end{array}\right)
\left(
\begin{array}{c}
p_{n}(0000)\\ 
p_{n}(1111)\end{array}\right)
\ee
$ n\ge 4$.

\noindent {\bf -Examples:} 

\noindent by direct computation $0\le n< 4$
\begin{eqnarray} p_0(0000)&=& {{\tilde\zeta(3)}\over 2^2}\nonumber \\
 i_1(0001)&=& {{\tilde\zeta(3)}\over 2^3 }\nonumber\\
  p_2(0000)&=&{-3+{\tilde\zeta(3)}\over 2^4}\nonumber  \\
 p_2(0011)&=&{1+{\tilde\zeta(3)}\over 2^4 }\quad {\rm known \cite{1}} \nonumber\\
 i_3(0001)&=&{-3+{\tilde\zeta(3)}\over 2^4 }\nonumber\\
 i_3(0111)&=&{1\over 2^2}\quad {\rm obvious}\nonumber
\end{eqnarray}
and $n=4$ (initial conditions)
\begin{eqnarray}
p_{4}(0000)&=&
  {-3^3+7{\tilde\zeta(3)}\over 2^6}\nonumber\\
 p_4(1111)&=&{53-3^2{\tilde\zeta(3)}\over 2^6}\nonumber
\end{eqnarray}
then the recurrence (\ref{rec}) gives for 
 $n>4$
\begin{eqnarray} p_6(0000)&=&{-37+3^2{\tilde\zeta(3)}\over 2^4}\nonumber\\
 p_6(1111)&=&3{67-(3)(5){\tilde\zeta(3)}\over 2^6}\nonumber\\
  p_{8}(0000)&=&{-(5)(19)(269)+(3^2)(7)(97){\tilde\zeta(3)}\over 2^{10}} \nonumber\\
 p_8(1111)&=&3{(13) (811)-(3^2)(5^2)(11){\tilde\zeta(3)}\over 2^{10}}\nonumber\\
  p_{10}(0000)&=&{-9304913+(3^2)(5^3)(11)(179){\tilde\zeta(3)}\over 
(2^{12})(5)} \nonumber\\
 p_{10}(1111)&=&3^4{(3)(11)(4139)-(5^3)(7)(37){\tilde\zeta(3)}\over 
(2^{12})(5)}\nonumber\\
  p_{12}(0000)&=&3^4{-(7)(19)(23909)+(5^3)(23)(263){\tilde\zeta(3)}\over 
(2^{12})(5)} \nonumber\\
 p_{12}(1111)&=&3^2{(43)(67)(11519)-(3^2)(5^3)(7^2)(11)(13){\tilde\zeta(3)}\over (2^{12})(5)}\nonumber
\end{eqnarray}

If one sets $n=2k$ and defines
$q_k(0)=p_{2k}(0000)/(2k)!$,
 $q_k(1)=p_{2k}(1111)/(2k)!$
 the recurrence  (\ref{rec}) becomes 
\be\label{recbis}
\left(
\begin{array}{c}
q_{k+1}(0)\\ 
q_{k+1}(1)\end{array}\right)
={1\over 2^4(2k+1)}
\left(
\begin{array}{cc}
(5k+2) & {-3k^2(k-1)/ (k+1)^2}\\  
 -3(k+2) & {k(k-1)(5k+8)/ (k+1)^2} 
\end{array}\right)
\left(
\begin{array}{c}
q_{k}(0)\\ 
q_{k}(1)\end{array}\right)
\ee
$k\ge 2$. 
In the asymptotic regime $k\to\infty$ the recurrence matrix 
\bea
{1\over 2^5}
\left(
\begin{array}{cc}
5 & -3\\  
-3 & 5 
\end{array}\right)
\nonumber\eea
 diagonalizes as 
\[
{1\over \sqrt{2}}
\left(
\begin{array}{cc}
1 & 1\\
-1 & 1 
\end{array}\right)
{1\over 2^5}\left(
\begin{array}{cc}
5 & -3\\  
 -3 & 5 
\end{array}\right)
{1\over \sqrt{2}}\left(
\begin{array}{cc}
1&-1\\
1&1 
\end{array}\right)
={1\over 2^4}
\left(
\begin{array}{cc}
1 & 0\\  
 0 & 4 
\end{array}\right)
\nonumber\]
with  eigenvalues $1/4$ and $1/16$.
In this convenient  diagonal basis 
(\ref{recbis}) becomes
\bea\label{toto}
\left(
\begin{array}{c}
q_{k+1}(1)+q_{k+1}(0)\\ 
q_{k+1}(1)-q_{k+1}(0)\end{array}\right)= \hskip 11.5truecm\nonumber \\
{1\over 2^4(2k+1)} 
\left(
\begin{array}{cc}
k-2+k(k-1)(k+4)/(k+1)^2 & -(k-2)+k(k-1)(k+4)/(k+1)^2\\  
 4(-(k+1)+k(k-1)/(k+1)) & 4(k+1+k(k-1)/(k+1))
\end{array}\right)\nonumber \\
\left(
\begin{array}{c}
q_{k}(1)+q_{k}(0)\\ 
q_{k}(1)-q_{k}(0)\end{array}\right) \hskip 10.5truecm
\nonumber\eea
$ k\ge 2$ with  by definition
$q_{k}(1)+q_{k}(0)>
q_{k}(1)-q_{k}(0)> 0$.

\noindent {\bf -Examples:}

\noindent
$k=2$ (initial conditions) 
\be\label{init1}
q_{2}(1)+q_{2}(0)={13-{\tilde\zeta(3)}\over (2^8)(3)}\ee 
\be\label{init2}
q_{2}(1)-q_{2}(0)={5-{\tilde\zeta(3)}\over (2^5)(3)}\ee 
$k>2$ 
\begin{eqnarray}
q_{3}(1)+q_{3}(0)&=&{53-(3)^2{\tilde\zeta(3)}\over (2^{10})(3^2)(5)}\nonumber\\
q_{3}(1)-q_{3}(0)&=&{349-(3)^4{\tilde\zeta(3)}\over (2^{10})(3^2)(5)}\nonumber\\
q_{4}(1)+q_{4}(0)&=&{3037-(3)^2(73){\tilde\zeta(3)}\over (2^{16})
(3^2)(5)(7)}\nonumber\\ 
q_{4}(1)-q_{4}(0)&=&{1787-(3)^2(47){\tilde\zeta(3)}\over
(2^{12}) (3^2)(5)(7) }\nonumber\\
q_{5}(1)+q_{5}(0)&=&{(439)(2003)-(3^2)(5^3)(181){\tilde\zeta(3)}\over
(2^{19})(3^4)(5^3)(7) }\nonumber\\
q_{5}(1)-q_{5}(0)&=&{(7)(73)(1993)-(3^2)(5^4)(43){\tilde\zeta(3)}\over
(2^{18})(3^4)(5^2)(7) }       \nonumber\\
q_{6}(1)+q_{6}(0)&=&{(2283583)-(3^2)(5^3)(479){\tilde\zeta(3)}\over
(2^{21})(3^3)(5^3)(7)(11) }       \nonumber\\
q_{6}(1)-q_{6}(0)&=&{(127)(1901)-(3^3)(5^3)(17){\tilde\zeta(3)}\over
(2^{14})(3^3)(5^3)(7)(11) }         \nonumber\\
q_{7}(1)+q_{7}(0)&=&{(53)(1708543)-(3)(5^3)(7^3)(167){\tilde\zeta(3)}\over
(2^{21})(3^2)(5^3)(7^3)(11)(13)}       \nonumber\\
q_{7}(1)-q_{7}(0)&=&{(13)(61485173)-(3)(5^3)(7^4)(211){\tilde\zeta(3)}\over
(2^{23})(3^2)(5^3)(7^2)(11)(13)}   \nonumber
\end{eqnarray}
Both  $q_{k}(1)+q_{k}(0)$ and $
q_{k}(1)-q_{k}(0)$ decrease (faster than $(1/16)^{k-2}$) as $(1/16)^{k-2}$
(when $k$ finite) when $k\to\infty$, i.e. their asymptotic behavior is
governed by the smallest of the eigenvalues. This is due to the
 initial  conditions fine tuning (\ref{init1}) and
(\ref{init2}), which defines a
 ${\tilde\zeta(3)}$ dependant
initial orientation in the 2 dimensional vector space leading 
to
this  particular asymptotic behavior.
This situation is analogous to the one encountered for establishing the
irrationality of $\zeta(3)$  (see \cite{3} for a review of
various demonstrations of the irrationality of $\zeta(3)$): in that case
 a second order recurrence relation (or
equivalently a family of integrals) with $\zeta(3)$ dependant initial
conditions such that an asymptotic convergence governed by the smallest of
eigenvalues is achieved. It is however manifest in the present case that both
$q_{k}(1)+q_{k}(0)$ and $ q_{k}(1)-q_{k}(0)$ have denominators which increase
too fast with $k$ to obtain a convergence allowing for an alternative proof of the irrationality of
$\zeta(3)$. This is still the case for any $k$-independant initial conditions
rotation in the plane. If one allows for more general $k$-dependant
combinations, note that if one redefines $ q_k(01)=p_{2k}(0011)/(2k)!$ then

\[ q_k(01)={q_k(1)+q_k(0)}-{(q_k(1)-q_k(0))/ k}\]
 if one redefines $j_k(0111)=i_{2k+1}(0111)/(2k+1)! $ then 
\[ j_k(0111)= {k-1\over 2(2k+1)}q_k(1)\] and finally 
if one redefines
$j_k(0001)= i_{2k+1}(0001)/(2k+1)! $ then 
\[ j_k(0001)= {k+1\over
2(2k+1)}q_k(0)\]
Particular $k$-dependant linear combinations of
${q_k(1)+q_k(0)}$ and $q_k(1)-q_k(0)$ follow as 
\[ j_k(0111)+j_k(0001)=
{k(q_k(1)+q_k(0))-(q_k(1)-q_k(0))\over 2(2k+1)}\le q_k(1)+q_k(0) \]
 \[
j_k(0111)-j_k(0001)= {k(q_k(1)-q_k(0))-(q_k(1)+q_k(0))\over 2(2k+1)}\le
q_k(1)-q_k(0)\]
with by definition $j_k(0111)-j_k(0001)>0$.
However both   $j_k(0111)-j_k(0001)$ and $j_k(0111)+j_k(0001)$  have  again 
denominators which increase too fast with $k$.

\section{Further Generalization: $\zeta(5),\zeta(7),\cdots $}
In view of obtaining $\zeta(5)$, a natural generalization of
 (\ref{11}) which would still involve Macdonald
functions would be, in the simplest case 
\[ \int_{a,b,c,d,e,f=0}^{\infty}\rmd a\,
\rmd b\, \rmd c\, \rmd d\,  \rmd e\, \rmd f\, {1\over abcde+bcdef +cdefa+defab+efabc+fabcd}
e^{-(a+b+c+d+e+f)}\] 
nothing but 
\[ 2^6\int_{0}^{\infty}{u\rmd u\,\over 2}
K_{0}(u)^6\] 
Numerics indicates however that this integral is not a linear
combination with rational coefficients of 1, $\zeta(5)$, $\zeta(3)$,
$\zeta(2)\zeta(3)$, \ldots. Introduce rather for $n\ge 2 $  the $2n $ 
integration variables
$a_i$ and $b_i$, with $0\le i\le n$, and consider  
\[
\int_0^{\infty}\prod_{i=1}^n \rmd a_i\,\rmd b_i\, \delta(1-\sum_{i=1}^n (a_i+b_i))\
\frac{P(a_1,b_1,\ldots,a_n,b_n)} {a_1b_1(a_2+b_2)\ldots(a_n+b_n)+{\rm cp}} \]
where cp means cyclic permutation over $1,2,...,n$ and ${P(a_1,b_1,\ldots,a_n,b_n)}$ is any given monomial of the variables
$a_i,b_i$. The case $n=2$ corresponds to the variables $a,b,c,d$ used up to
now. 
This is a family of integrals with root
\[
I_{n}=\int_0^{\infty}\prod_{i=1}^n \rmd a_i\,\rmd b_i\, \delta(1-\sum_{i=1}^n
(a_i+b_i))\ \frac{1} {a_1b_1(a_2+b_2)\ldots(a_n+b_n)+{\rm cp}} \]
Change variables 
\[ u_i=a_i+b_i\qquad v_i=\frac{a_i-b_i}{a_i+b_i} \]
to get 
\[
I_{n}=4\int_0^{\infty}\prod_{i=1}^n \rmd u_i\, \delta(1-\sum_{i=1}^n
u_i) \int_0^1\prod_i^n \rmd v_i\, \frac{1} {(1-v_1^2)u_1+\ldots+(1-v_n^2)u_n} \]
In order to integrate over $u_1\ldots u_n$, follow the procedure used above:
Laplace transform the $\delta$ measure and exponentiate the denominator 
\[
I_{n}=\frac{4}{\Gamma(n-1)} \int_0^1\prod_i^n
\rmd v_i\,\int_0^{\infty}\rmd t\,\prod_i^n \rmd u_i \, \exp\left(-\sum_{i=1}^n
u_i(1+t(1-v_i^2))\right) \] 
Integration over $u_i$ and $v_i$ leads to
\[
I_{n}=\frac{4}{2^n\Gamma(n-1)} \int_0^{\infty}\rmd t\, \frac{1}
{(\sqrt{(1+t)t})\,^n}
\left[\log\left(\frac{\sqrt{\frac{1+t}{t}}+1} {\sqrt{\frac{1+t}{t}}-1}\right)
\right]^n \]
 which can be rewritten\footnote{or as \[ I_{n}=\frac{2}{\Gamma(n-1)}
\int_0^{\infty} \rmd y\,{y}({y\over \sh y})^{n-1}\]} as 
\[ I_{n}=\frac{2^n}{\Gamma(n-1)}
\int_0^1 \frac{\rmd y\,}{y}\left(\frac{y}{1-y^2}\right)^{n-1} (\log {1\over y})^n \] 
Using
\[ \int_0^1 \rmd y\, y^p(\log {1\over y})^n = \frac{\Gamma(n+1)}{(p+1)^{n+1}} \]
leads to 
\be\label{final} I_{n}=2^n\frac{\Gamma(n+1)}{\Gamma(n-1)^2}
\sum_{p=0}^{\infty}\frac{(p+1)(p+2)\ldots(p+n-2)}{(n+2p-1)^{n+1}} \ee

\noindent{\bf -Examples:}
\begin{eqnarray}
I_2 &=& {7}\zeta(3)\nonumber\\[0.2cm]
I_3 &=& {3}\zeta(3)\nonumber\\[0.2cm]
I_4 &=& {3}({7}\zeta(3)-\frac{31}{2^2}\zeta(5))\nonumber\\[0.2cm]
I_5 &=& \frac{5}{3}(\zeta(3)-\zeta(5))\nonumber\\[0.2cm]
I_6 &=& \frac{5}{2^3}({7}\zeta(3)
-\frac{(5)(31)}{2}\zeta(5)+\frac{(3^2)(127)}{2^4}\zeta(7))\nonumber
\end{eqnarray}
Only $\zeta({\rm odd})$ does  appear in (\ref{final}) as can be seen by 
distinguishing between $n$ odd and $n$ even
\begin{eqnarray}
I_{n}&=&\frac{4\Gamma(n+1)}{\Gamma(n-1)^2}
\sum_{p=0}^{\infty}\frac{\prod_{q=1}^{(n-2)/2}\bigl((n+2p-1)^2-(2q-1)^2\bigr)}{(n+2p-1)^{n+1}}\qquad
n \quad {\rm even}\nonumber   \\[0.2cm]
I_{n}&=&\frac{4\Gamma(n+1)}{\Gamma(n-1)^2}
\sum_{p=0}^{\infty}\frac{\prod_{q=1}^{(n-3)/2}\bigl((n+2p-1)^2-(2q)^2\bigr)}{(n+2p-1)^n}\qquad
\quad \quad n\quad {\rm odd} 
\nonumber\end{eqnarray}

\textbf{Acknowledgments:} One of us (S.O.) would like to thank J. Desbois, 
C. Jacquemin, J. Myrheim, C. Schmit and P. Zinn-Justin
for useful conversations.

\end{document}